\documentclass[prd,twocolumn,a4paper,superscriptaddress,floatfix]{revtex4}
\usepackage{graphicx}
\usepackage{bbm}
\usepackage[all]{xy}
\usepackage{amsmath}
\usepackage{amssymb}
\usepackage{epstopdf}

\def\be{\begin{equation}}
\def\ee{\end{equation}}
\newcommand{\bq}{\begin{eqnarray}}
\newcommand{\eq}{\end{eqnarray}}
\newcommand{\bes}{\begin{subequations}}
\newcommand{\ees}{\end{subequations}}

\def\ben{\begin{eqnarray}}
\def\een{\end{eqnarray}}
\def\ba{\begin{array}}
\def\ea{\end{array}}

\begin{document}
\newcommand{\half}{{\textstyle\frac{1}{2}}}
\allowdisplaybreaks[3]
\def\a{\alpha}
\def\b{\beta}
\def\g{\gamma}\def\G{\Gamma}
\def\d{\delta}\def\D{\Delta}
\def\ep{\epsilon}
\def\et{\eta}
\def\z{\zeta}
\def\t{\theta}\def\T{\Theta}
\def\l{\lambda}\def\L{\Lambda}
\def\m{\mu}
\def\f{\phi}\def\F{\Phi}
\def\n{\nu}
\def\p{\psi}\def\P{\Psi}
\def\r{\rho}
\def\s{\sigma}\def\S{\Sigma}
\def\ta{\tau}
\def\x{\chi}
\def\o{\omega}\def\O{\Omega}
\def\k{\kappa}
\def\pa {\partial}
\def\ov{\over}
\def\br{\\}
\def\ud{\underline}

\newcommand\lsim{\mathrel{\rlap{\lower4pt\hbox{\hskip1pt$\sim$}}
    \raise1pt\hbox{$<$}}}
\newcommand\gsim{\mathrel{\rlap{\lower4pt\hbox{\hskip1pt$\sim$}}
    \raise1pt\hbox{$>$}}}
\newcommand\esim{\mathrel{\rlap{\raise2pt\hbox{\hskip0pt$\sim$}}
    \lower1pt\hbox{$-$}}}
\newcommand{\dpar}[2]{\frac{\partial #1}{\partial #2}}
\newcommand{\sdp}[2]{\frac{\partial ^2 #1}{\partial #2 ^2}}
\newcommand{\dtot}[2]{\frac{d #1}{d #2}}
\newcommand{\sdt}[2]{\frac{d ^2 #1}{d #2 ^2}}    

\title{Scaling laws for weakly interacting cosmic (super)string and $p$-brane networks}

\author{P.P. Avelino}
\email[Electronic address: ]{ppavelin@fc.up.pt}
\affiliation{Centro de Astrof\'{\i}sica da Universidade do Porto, Rua das Estrelas, 4150-762 Porto, Portugal}
\affiliation{Departamento de F\'{\i}sica da Faculdade de Ci\^encias
da Universidade do Porto, Rua do Campo Alegre 687, 4169-007 Porto, Portugal}
\author{L. Sousa}
\email[Electronic address: ]{Lara.Sousa@astro.up.pt}
\affiliation{Centro de Astrof\'{\i}sica da Universidade do Porto, Rua das Estrelas, 4150-762 Porto, Portugal}

\begin{abstract}

In this paper we find new scaling laws for the evolution of $p$-brane networks in $N+1$-dimensional Friedmann-Robertson-Walker universes in the weakly-interacting limit, giving particular emphasis to the case of cosmic superstrings ($p=1$) living in a universe with three spatial dimensions ($N=3$). In particular, we show that, during the radiation era, the root-mean-square velocity is ${\bar v} =1/{\sqrt 2}$ and the characteristic length of non-interacting cosmic string networks scales as $L \propto a^{3/2}$ ($a$ is the scale factor), thus leading to string domination even when gravitational backreaction is taken into account. We demonstrate, however, that a small non-vanishing constant loop chopping efficiency parameter $\tilde c$ leads to a linear scaling solution with constant $L H  \ll 1$ ($H$ is the Hubble parameter) and ${\bar v} \sim 1/{\sqrt 2}$ in the radiation era, which may allow for a cosmologically relevant cosmic string role even in the case of light strings. We also determine the impact that the radiation-matter transition has on the dynamics of weakly interacting cosmic superstring networks.

\end{abstract} 
\maketitle

\section{Introduction}

Cosmic strings are usually regarded as benign objects with potentially interesting cosmological consequences. Ordinary cosmic strings have an efficient energy-loss mechanism: once two strings collide, they exchange partners and intercommute. This results in the creation of string loops, which eventually decay radiatively. The corresponding energy loss aids the attainment of a linear scaling regime, during which the average energy density of the string network remains a constant fraction of the background energy density \cite{Kibble:1976sj,Kibble:1980mv}. The existence of this regime has not only been confirmed numerically \cite{Bennett:1987vf,Albrecht:1989mk,Allen:1990tv}, but was also established using semi-analytical models \cite{Copeland:1991kz,Vincent:1996rb,Martins:1996jp,Martins:2000cs}.

Recent developments in string theory indicate that fundamental strings (F-strings) or 1-dimensional Dirichlet branes (D-strings) may grow to macroscopic scales and play the role of cosmic strings \cite{Witten:1985fp,Majumdar:2002hy,Jones:2003da,Copeland:2003bj}. These cosmic superstrings may have an intercommuting probability significantly smaller than unity \cite{Jackson:2004zg}, and thus cosmic superstring networks may be weakly interacting. It is then crucial to understand the effect that a significantly less efficient energy-loss mechanism might have on the late-time evolution of the network, and to determine the conditions under which it might prevent the string network from dominating the energy density of the universe.

In this article, we revisit the question of whether or not weakly interacting string networks are able to attain linear scaling regimes. We start by reviewing, in Sec. \ref{dyn}, the dynamics of $p$-branes and the Velocity-Dependent One-Scale (VOS) model for $p$-brane networks. In Sec. \ref{sca}, we obtain the frictionless scaling laws for weakly interacting $p$-brane networks. In Sec. \ref{string}, we consider the particular case of cosmic superstrings in $3+1$-dimensional Friedmann-Robertson-Walker (FRW) backgrounds, and determine the impact that gravitational backreaction has on the dynamics of the network. We then conclude in Sec \ref{conc}.

\section{$p$-brane network dynamics}\label{dyn}

In the zero-thickness limit, the world-history in spacetime of a featureless $p$-brane may be represented by
\be
x^{\mu}=x^{\mu}(u^{{\tilde \nu}})\,,
\ee
where $u^{\tilde \nu}$ with ${\tilde \nu}=0,1,...,p$ are the coordinates parameterizing the $(p+1)$-dimensional worldsheet swept by the $p$-brane, $u^0$ is a timelike parameter and $u^{\tilde i}$ are spacelike parameters. The action of a thin and featureless $p$-brane is described by the Nambu-Goto action
\be
S=-\sigma_p \int d^{p+1}u\sqrt{\left|{\tilde g}\right|}\,,
\label{nambugoto}
\ee
where $\sigma_p$ is the (constant) $p$-brane mass per unit $p$-dimensional area, ${\tilde g}=\det({\tilde g}_{{\tilde \mu} {\tilde \nu}})$, ${\tilde g}_{{\tilde \mu} {\tilde \nu}}=g_{\alpha \beta} x^{\alpha}_{,{\tilde \mu}}x^{\beta}_{,{\tilde \nu}}$ is the worldsheet metric, $x^{\alpha}_{,{\tilde \mu}}=\partial x^{\alpha}/\partial u^{\tilde \mu}$ and $g_{\mu\nu}$ is the metric tensor. 

In a $(N+1)$-dimensional flat Friedmann-Robertson-Walker Universe, the line element is given by
\be
ds^2=a^2(\eta)\left(d^2\eta-{d\mathbf{x}}\cdot d{\mathbf{x}}\right)\,,
\ee
where $a$ represents the cosmological scale factor, $t$ and $\eta=\int dt/a$ are respectively the physical and conformal time and ${\bf x}$ is a $N$-vector whose components are comoving cartesian coordinates. By varying the action in Eq. (\ref{nambugoto}) with respect to $x^\mu$ and imposing temporal-transverse gauge conditions, 
\bq
u^0&=&\eta\,, \\
\dot{\mathbf{x}}\cdot\mathbf{x}_{,{\tilde i}}&=&0\,, \label{gauge1}
\eq
one obtains the equation of motion \cite{Sousa:2011iu}
\bq
\ddot{\mathbf{x}} & + & \left(p+1\right) \mathcal{H}(1-\dot{\mathbf{x}}^2)\dot{\mathbf{x}} =\nonumber\\
& =& \epsilon^{-1} \sum_{{\tilde i}=1}^p\left[\frac{\mathbf{x}_{,{\tilde i}}}{\epsilon}\Pi_{{\tilde j}\neq {\tilde i}}(\mathbf{x}_{,{\tilde j}})^2\right]_{,{\tilde i}}\,,\label{ngeq}\\
\dot{\epsilon} & = & -\left(p+1\right)\mathcal{H}\epsilon \dot{\mathbf{x}}^2\,.
\eq
Here
\be
\epsilon=\left(\frac{(\mathbf{x}_{,1})^2\cdots (\mathbf{x}_{,p})^2}{1-\dot{\mathbf{x}}^2}\right)^{\frac{1}{2}} \label{epsilon}\,,
\ee
a dot represents a derivative with respect to conformal time, $\mathbf{x}_{,{\tilde i}} = \partial \mathbf{x}/ \partial u^{\tilde i}$, $u^{\tilde i}$ are orthogonal spacelike worldsheet coordinates (with ${\tilde i}=1,\cdots,p$), and $\mathcal{H}=\dot{a}/a$. Given the  gauge conditions described in Eq. (\ref{gauge1}), $\dot{\bf x}$ represents the physical velocity of the $p$-brane and it is perpendicular to the brane itself.

The total energy $E$ and root-mean-square (RMS) velocity $\bar v$ of a $p$-brane network are defined respectively by
\bq
E&=&\sigma_p a^p \int \epsilon d^p u\,, \label{Edef}\\
{\bar v}^2 &=& \frac{\int {\dot{\bf x}}^2 \epsilon d^p u}{\int \epsilon d^p u}\label{vdef}\,,
\eq
so that
\be
{\dot E}=(p+1)\mathcal{H} E \left(\frac{p}{p+1}-{\bar v}^2\right)\,.
\ee
For very small $p$-branes it is a good approximation to consider that the expansion of the universe has, on average, no impact on the total energy, so that the averages over a sufficiently long time of the total energy $\langle E \rangle_t$ and RMS velocity $\langle {\bar v}^2\rangle_t =p/(p+1)$ are constant.

In the case of a statistically homogeneous $p$-brane network, the average $p$-brane energy density may be defined as $\rho_p=E/V$, with $V \propto a^N$. It is then straightforward to show that
\be
{{\dot \rho}_p}+\mathcal{H}{\rho_p}\left[D+\left(p+1\right){\bar v}^2\right]=0\,,
\label{vos-den}
\ee
where $D=N-p$. By defining the characteristic length, $L$, of the network as
\be
\rho_p=\frac{\sigma_p}{L^D}\,,
\label{Ldef}
\ee
this equation may be rewritten as
\be
\frac{dL}{dt}=HL\left(1+(p+1)\frac{{\bar v}^2}{D}\right)\,,
\label{vos-Lpre}
\ee
where $H=\mathcal{H}/a$ is the Hubble parameter.

\subsection{VOS model}

A unified VOS model for the dynamics of $p$-brane networks in $(N+1)$-dimensional FRW universes was derived in \cite{Avelino:2011ev,Sousa:2011ew}. According to this model, the cosmological evolution of a statistically homogeneous $p$-brane network may be described by the following equations
\bq
\frac{d{\bar v}}{dt}&+&\left(1-{\bar v}^2\right)\left[\frac{{\bar v}}{\ell_d}-\frac{k}{L}\right]=0\,, \label{VOS_v}\\
\frac{dL}{dt}&=&HL+\frac{L}{D \ell_d}{\bar v}^2+ \frac{{\tilde c}}{D}{\bar v}\,,
\label{vos-L}
\eq
where $\ell_d^{-1}=(p+1)H$ is the damping lengthscale, $k$ is a dimensionless curvature parameter and the assumption that $\left<v^4\right>={\bar v}^4$ was made (see \cite{Martins:1996jp}). Eq. (\ref{vos-L}) is identical to Eq. (\ref{vos-Lpre}) apart from the energy loss term associated with $p$-brane reconnection (last term) while Eq.  (\ref{VOS_v}) has been rigorously derived using Eqs.  (\ref{ngeq}),  (\ref{epsilon}) and  (\ref{vdef}) in \cite{Avelino:2011ev,Sousa:2011ew,Sousa:2011iu} (the curvature parameter $k$ is also rigorously defined therein). A frictional force --- caused by the interaction of the branes with ultrarelativistic particles or other frictional sources --- may also be included in Eq. (\ref{VOS_v}), by introducing a friction lengthscale $\ell_f$ so that $\ell_d^{-1}=(p+1)H+\ell_f^{-1}$. For simplicity, we will not consider it in the present paper and consequently we shall take $\ell_f=\infty$ (see \cite{Sousa:2011ew} for a detailed discussion of friction dominated regimes). 


\section{Scaling regimes\label{sca}}

In a FRW universe with a decelerating power-law expansion --- $a \propto t^\beta$ with $0\le \beta<1$ --- Eqs. (\ref{VOS_v}) and (\ref{vos-L}) admit linear attractor solutions of the form
\be
L=\xi t \qquad \mbox{and} \qquad \bar v=\mbox{constant}\,,
\label{linearscalinga}
\ee
with 
\bq
\xi &=& \sqrt{\left|\frac{k(k+{\tilde c})}{\beta (1-\beta)D(p+1)}\right|}\label{linearscaling}\,,\\
{\bar v} &=& \sqrt{\frac{(1-\beta)kD}{\beta(k+{\tilde c})(p+1)}}\label{linearscaling1}\,.
\eq
The existence of an energy-loss mechanism helps the attainment of a linear scaling solution but that is not a necessary condition. It is a common misconception that, if ${\tilde c}=0$, these solutions are possible for all $\beta>D/(N+1)$ (or, equivalently ${\bar v}^2 < 1$ in Eq. (\ref{linearscaling1})). However, in expanding backgrounds, it is reasonable to expect the RMS velocity to be smaller than it would be in Minkowski space. One should then expect $k({\bar v})$ to vanish for ${\bar v}^2=p/(p+1)$ and the scaling velocity not to exceed this value. This, in turn, implies that the linear scaling solutions are, in the absence of energy-loss mechanisms due to $p$-brane reconnection, attainable only for $\beta\ge D/N$. In a $3+1$ dimensional FRW universe, the threshold is then $\beta=2/3$ (rather than $\beta=1/2$) for cosmic strings, and $\beta=1/3$ (instead of $\beta=1/4$) for domain walls.

\subsection{Non-interacting $p$-branes\label{non-int}}

For ${\tilde c}=0$ and $0 \le \beta \le D/N$, $p$-brane networks are expected to evolve towards a solution with ${\bar v}^2 = p/(p+1)$. The equation-of-state parameter of a brane gas is given by \cite{Boehm:2002bm,Sousa:2011iu}
\be
w_p=\frac{{\mathcal P_p}}{\rho_p}=\frac{1}{N}\left[\left(p+1\right){\bar v}^2-p\right]\,.
\label{eos}
\ee
where ${\mathcal P_p}$ is the average brane pressure. One may then conclude that, for $0 \le \beta \le D/N$ (and ${\bar v}^2 = p/(p+1)$), the network can be described, on scales much larger than $L$, as a quasi-homogeneous matter background with $w_p=0$.


If  ${\tilde c}=0$, Eq. (\ref{vos-L}) may be written as
\be
\frac{dL}{d \ln a}=L\left(1+(p+1)\frac{{\bar v}^2}{D}\right)\,.
\label{vos-L1}
\ee
We then have that, if  $0\le \beta \le D/N$ (and ${\bar v}^2=p/(p+1)$), the characteristic lengthscale of the network scales as $L\propto a^{N/D}$. Hence, in this case, the characteristic length evolves asymptotically as
\be
L\propto t^\alpha\,,\quad\mbox{with}\quad \alpha=\beta\frac{N}{D} \le 1\,.
\label{L-small}
\ee

\subsection{Weakly interacting case\label{weak}}

The picture changes significantly if we consider a constant non-vanishing energy loss parameter $0 < {\tilde c} \ll 1$. This case is relevant not only for the study of cosmic superstring networks --- which might have a small intercommuting probability --- but also for the study of $p$-brane networks in $N$-dimensional backgrounds. In particular, if $p<(N-1)/2$, $p$-branes are likely to miss each other \cite{Sousa:2011ew}.

As previously discussed, the curvature parameter should vanish and change sign at ${\bar v}^2=p/(p+1)$.  Let us define the parameter $k({\bar v})$ as
\be
k({\bar v})=A f ({\bar v})\,,
\label{kv}
\ee
where $A$ is a constant and
\be
f({\bar v})=\left({\frac{p}{p+1}}-{\bar v}^2\right)\,.
\label{kv1}
\ee
The curvature parameter defined Eqs. (\ref{kv}) and (\ref{kv1}) may be considered a good approximation if $f \ll 1$. In this case, for ${\tilde c} \ll 1$ and $\beta < D/N$, a linear scaling solution (see Eq. (\ref{linearscalinga})) with
\be
\xi = \frac{{\tilde c}p^{1/2}}{(p+1)^{1/2}(D-\beta N)}\,,  \quad f = \frac{\beta p^{1/2}(p+1)^{1/2}}{A}\xi 
\label{lins}
\ee
may be found up to first order in ${\tilde c}$ and $f$. Note that, in the ${\beta \to D/N}$ limit, one needs to use Eqs. (\ref{linearscalinga})-(\ref{linearscaling1}) in combination with Eqs.  (\ref{kv}) and (\ref{kv1}) in order to find the scaling solutions.

\section{Cosmic strings\label{string}}

Let us consider the more interesting case of cosmic superstring networks in a 3+1 dimensional FRW background (with $N=3$, $p=1$, and $\mu \equiv \sigma_1$). The generalization to other values of $p$ and $N$ is trivial. It follows from Eq. (\ref{L-small}) that during the radiation era, with $\beta=1/2$, one has $L \propto a^{3/2} \propto t^{3/4}$, in the absence of any energy-loss mechanisms. Therefore, the string energy density $\rho_\mu=\rho_1=\mu/L^2$ is proportional to $a^{-3}$ and tends to dominate the energy density of the universe during the radiation era.

\subsection{Gravitational Backreaction\label{grav}}

The effect of gravitational backreaction on the networks' dynamics may be taken into account by adding the following term to the right-hand side of Eq. (\ref{vos-L})
\be
\left(\frac{dL}{dt}\right)_{\rm gr}=4 {\tilde \Gamma} G \mu {\bar v}^6\,,
\label{vosgr}
\ee
with ${\tilde \Gamma} \sim 65$ \cite{Vilenkin:1981bx,Quashnock:1990wv,Allen:1991bk}. In this case, although a scaling solution given by Eq. (\ref{linearscalinga}) with
\be
\xi = 2 {\tilde \Gamma} G \mu \,,  \quad {\bar v}^2 = \frac{1}{2}\,,
\ee
would be mathematically correct, it would imply
\be
\left(\frac{\rho_\mu}{\rho_b}\right)_{\rm rad}=\frac{8\pi}{{\tilde \Gamma}^2G\mu}\label{radcons}\,,
\ee
where $\rho_b$ is the background density during the radiation dominated era ($\rho_b=1/(32 \pi G t^2)$). However, this possibility is completely excluded by observations: the cosmic string tension is constrained to be $G\mu \lesssim 10^ {-7}$ \cite{Avelino:2003nn,Dunkley:2010ge,Urrestilla:2011gr,Dvorkin:2011aj,Sanidas:2012ee}. Thus, for any reasonable value of $G \mu$, the right hand side of Eq. (\ref{radcons}) is much larger than unity. The inclusion of the effects of gravitational backreaction does not prevent cosmic strings from dominating the energy density of the universe.

On the other hand, one might consider a constant non-zero energy-loss parameter such that $G\mu \ll{\tilde c} \ll 1$ and take the momentum parameter proposed in  \cite{Martins:2000cs}
\be
k({\bar v})=\frac{2{\sqrt 2}}{\pi}(1-{\bar v}^2)(1+2 {\sqrt 2} {\bar v}^3)\frac{1-8{\bar v}^6}{1+8{\bar v}^6}\,.
\label{kstring}
\ee
Then, during the radiation era, the VOS equations admit a linear scaling solution as in Eq. (\ref{linearscalinga}), with
\be
\xi=\sqrt{2}{\tilde c}\,,\quad {\bar v}=\frac{1}{\sqrt 2}-\delta\,,
\label{scaling-rad}
\ee
and
\be
\delta=\frac{\tilde c}{B}\,,
\label{deviation} 
\ee
up to first order in $\tilde{c}$ (here $B=12/\pi$). Note that during this regime $\bar v$ is not exactly $1/\sqrt{2}$. The deviation $\delta$ was be estimated by proceeding in a similar fashion as in Sec. \ref{weak} and by noting that, up to first order in $(1-\sqrt{2}{\bar v})$, Eq. (\ref{kstring}) may be written as 
\be
k({\bar v}) = B \left(1/\sqrt{2}-{\bar v}\right)\,.
\ee

During this regime, one has
\be
\left(\frac{\rho_\mu}{\rho_b}\right)_{\rm rad}=16 \pi \frac{G\mu}{{\tilde c}^2}\,,
\ee
which means that cosmic strings could make a fairly large contribution to a non-clustering matter background during the radiation era, even for low values of $G \mu$, as long as ${\tilde c}$ is small enough. For instance, if ${\tilde c}=10^{-2}$ and ${G \mu}=10^{-12}$ then $\left({\rho_\mu}/{\rho_b}\right)_{\rm rad} = 5 \times 10^{-7}$. On the other hand, if ${\tilde c}=10^{-3}$ and ${G \mu}=10^{-9}$ then $\left({\rho_\mu}/{\rho_b}\right)_{\rm rad} = 5 \times 10^{-2}$. Note that ${\tilde c}$ is expected to be proportional to the string intercommuting probability which may be significantly smaller than unity. According to \cite{Jackson:2004zg} values of ${\tilde c}$ within the range $[10^{-4},10^{-1}]$ appear to be acceptable (${\tilde c} \sim 0.23$ if the string intercommuting probability is equal to unity \cite{Martins:2000cs}).

\subsection{Radiation-Matter Transition\label{tran}}

\begin{figure}
\centering
\includegraphics[width=3.2in]{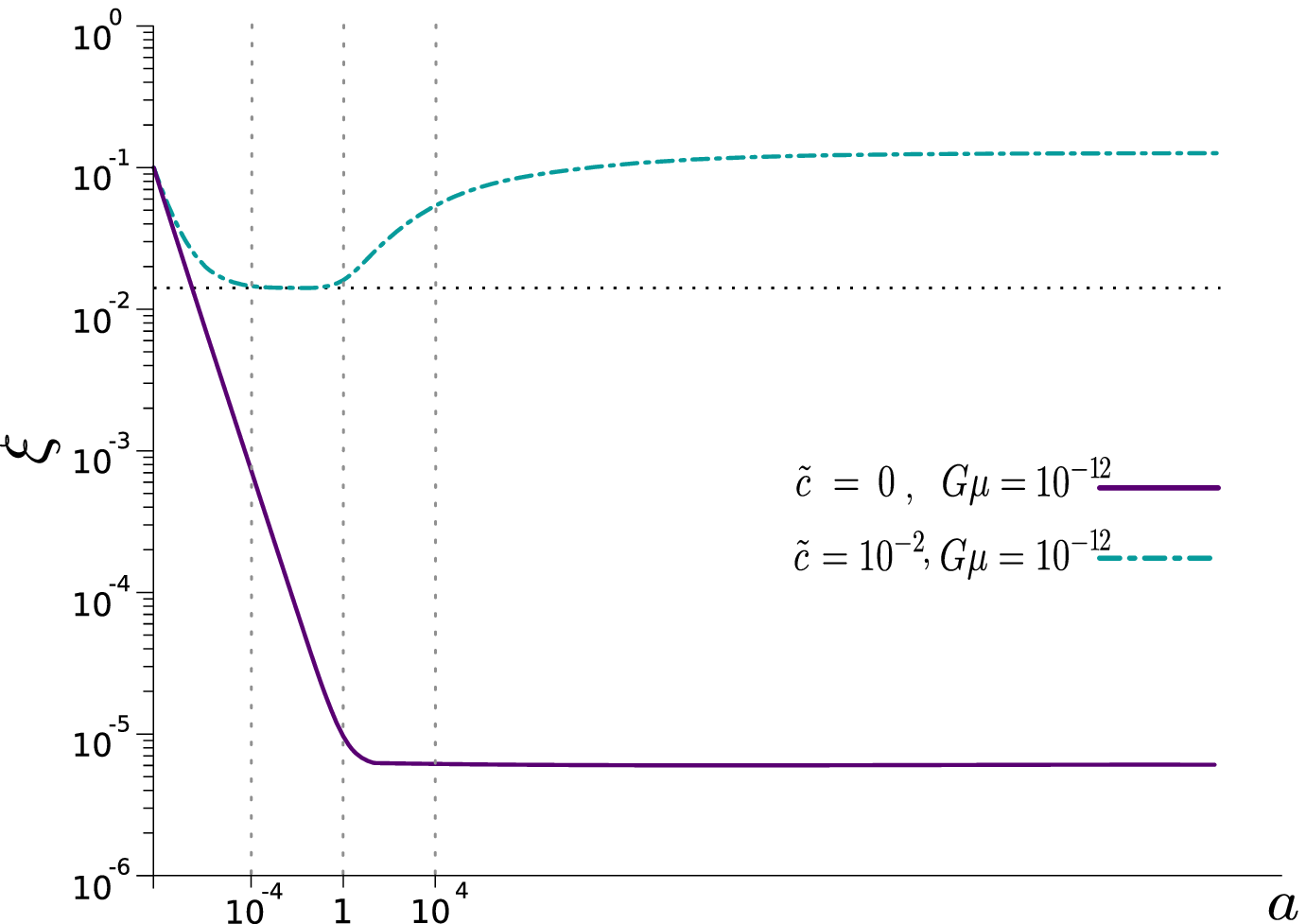}
\caption{Evolution of $\xi=L/t$ for a network of cosmic strings in a flat $3+1$-dimensional FRW universe with $G\mu=10^ {-12}$ and $\tilde{c}=0$ (solid (purple) line) or $\tilde{c}=10^ {-2}$ (dash-dotted (blue) line) as a function of the scale factor , $a$. The (black) dotted horizontal line indicates the expected value of $\xi$ during the radiation era (see Eq. (\ref{scaling-rad})). The value of the scale factor at radiation-matter equality is normalized to unity.}
\label{tran-xi}
\includegraphics[width=3.6in]{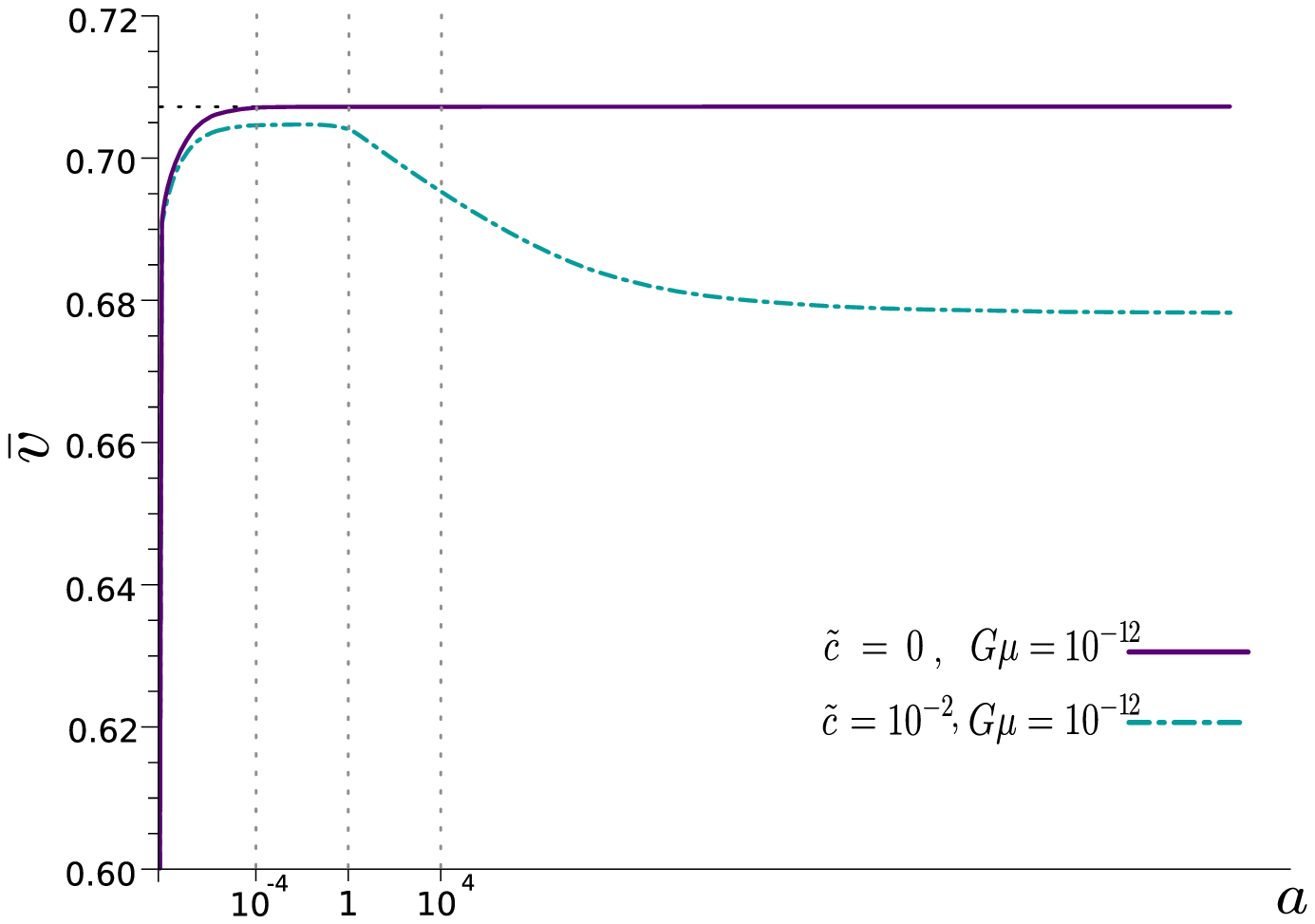}
\caption{Evolution of the RMS velocity, $\bar{v}$, for a network of cosmic strings in a flat $3+1$-dimensional FRW universe with $G\mu=10^ {-12}$ and $\tilde{c}=0$ (solid (purple) line) or $\tilde{c}=10^ {-2}$ (dash-dotted (blue) line) as a function of the scale factor, $a$. The (black) dotted horizontal line corresponds to $\bar{v}=1/\sqrt{2}$.}
\label{tran-v}
\end{figure}

Let us consider a flat FRW universe containing only matter and radiation, thus neglecting the dynamical effects associated with the recent acceleration of the expansion of the universe. Using the VOS equations (Eqs. (\ref{VOS_v}) and (\ref{vos-L})), it is possible to find numerical solutions for the evolution of non-interacting and weakly interacting cosmic string networks throughout the radiation-matter transition. The results are illustrated in Figs. \ref{tran-xi} and \ref{tran-v}, which show the evolution of $\xi=L/t$ and $\bar{v}$ for a network of cosmic strings in a flat $3+1$-dimensional FRW universe with $G\mu=10^ {-12}$ and $\tilde{c}=0$ (solid (purple) line) or $\tilde{c}=10^ {-2}$ (dash-dotted (blue) line) as a function of the scale factor, $a$ (the value of the scale factor at radiation-matter equality is normalized to unity). 

Figs. \ref{tran-xi} and \ref{tran-v} show that, as previously discussed, cosmic string networks are unable to attain a linear scaling regime during the radiation era if ${\tilde c}=0$. Instead, the RMS velocity and the characteristic length of the network evolve rather quickly towards a scaling regime with ${\bar v}^2=1/2$ and $\xi = L/t \propto t^{-1/4} \propto a^{-1/2}$. During the radiation-matter transition the rate of change of $\xi$ decreases and, eventually, $\xi$ tends to a constant. The network attains a linear scaling regime, with constant $\xi$, only when the universe becomes matter-dominated. Note, however, that the scaling value of $\xi$ is not defined in this case: it is simply the value of $\xi$ at the onset of matter domination and consequently it depends on the initial conditions.

On the other hand, if energy loss due to loop formation is taken into consideration (with $G\mu\ll\tilde{c}\ll 1$), then cosmic string networks may undergo two separate linear scaling regimes. If the string-forming phase transition occurs deep enough in the radiation era, the network is able to attain the regime in Eq. (\ref{scaling-rad}). As Fig. \ref{tran-v} clearly illustrates the value of $\bar v$ is fairly close to $\bar v \sim 1/\sqrt{2}$ in the radiation era, but there is a slight deviation. In this particular case, the value of $\delta$ is in good the agreement with the theoretical prediction in Eq. (\ref{deviation}): the relative difference is smaller than $1\%$. As the universe enters the matter dominated era, the network undergoes a phase during which $\xi$ grows (quasi-logarithmically) and $\bar{v}$ decreases as time progresses, until it reaches the linear scaling solution defined in Eqs. (\ref{linearscalinga})-(\ref{linearscaling1}) with $k({\bar v})$ given in Eq. (\ref{kstring}). Note that, the smaller the value of $\tilde{c}$, the longer it takes the network to attain the scale invariant regime in the matter-dominated era. For weakly interacting networks the transition phase may indeed be long lasting.

\section{Conclusions \label{conc}}

In this paper we determined the macroscopic evolution of $p$-brane networks in $N+1$-dimensional FRW universes with a decelerating power-law expansion $a \propto t^\beta$ ($0 \le \beta<1$). We found that, for $\beta < D/N$, a linear scaling solution with $L \propto t$ is possible only if the energy loss term due to $p$-brane reconnection and decay (parametrized by ${\tilde c}$) is non-zero. We have shown that, in this case, if $\tilde c=0$ then $L  \propto a^{N/D} \propto t^{\beta N/D}$, eventually leading to a $p$-brane dominated universe. We have further shown that if $\beta < D/N$ and the $p$-brane network is weakly interacting (with a small but non-zero ${\tilde c}$) the network then evolves towards a linear scaling solution with ${\bar v} \sim p/(p+1)$ and $L H \ll 1$. The corresponding equation-of-state parameter is $w_p\sim0$ thus generating a quasi-homogeneous matter background on scales much larger than $L$.

Our results have profound implications for the dynamics of cosmic superstring networks in expanding $3+1$-dimensional FRW universes. If ${\tilde c}=0$ then the root-mean-square velocity tends to ${\bar v} =1/{\sqrt 2}$ and the characteristic scale of non-interacting cosmic string networks scales as $L \propto a^{3/2}$ during the radiation era. We have shown that this leads to string domination even when gravitational backreaction is taken into account. On the other hand, a small non-zero constant loop chopping efficiency parameter $\tilde c$ leads to a linear scaling solution with constant $L H  \ll 1$ ($H$ is the Hubble parameter) and ${\bar v}\sim 1/{\sqrt 2}$ in the radiation era, thus generating a quasi-homogeneous matter background on scales much larger than the characteristic length $L$ of the network. We have shown that, in the scaling regime, the ratio between the average string energy density and the total background density is proportional to $G \mu/{\tilde c}^2$ which may allow for a cosmologically relevant cosmic string role even for light strings. We have also found that the transition from the radiation to the matter dominated era has a major impact on the dynamics of weakly interacting cosmic superstring networks.

\begin{acknowledgments}

This work is partially supported by FCT-Portugal through project CERN/FP/116358/2010 and through the grant SFRH/BPD/76324/2011.

\end{acknowledgments}


\bibliography{scaling}

\end{document}